\documentstyle[eqsecnum,aps]{revtex}

\begin{document}

\draft

\title{String cosmology coupled to Weyl-integrable geometry}

\author{ Israel Quiros\thanks{israel@uclv.etecsa.cu}}
\address{ Departamento de Fisica. Universidad Central de Las Villas. Santa Clara. CP: 54830 Villa Clara. Cuba }

\date{\today}

\maketitle

\begin{abstract}

The requirement that the laws of physics must be invariant under point-dependent transformations of the units of length, time, and mass is used as a selection principle while studying different generic effective theories of gravity. Thereof theories with non-minimal coupling of the dilaton both to the curvature and to the Lagrangian of the matter fields seem to represent the most viable low-energy [and low-curvature] description of gravity. Consequently, the cosmological singularity problem is treated within the context of string cosmology with non-minimal coupling of the dilaton to a barotropic gas of solitonic p-brane. The results obtained are to be interpreted on the grounds of Weyl-integrable geometry. The implications of these results for the Mach's principle are briefly discussed.

\end{abstract}


\section{Introduction}

In this paper the cosmological singularity problem will be treated from the point of view that is based upon the conformal transfromation technique. This technique has been often used in scalar-tensor theories, non-linear theories of gravity, Kaluza-Klein theory and string cosmology [among others] to put the field equations in a form that is mathematically more easy to handle\cite{fgn}.

The basic idea to be discussed here is linked with a geometrical interpretation of the conformal transformations of the spacetime metric that is due to Dicke. Under a conformal transformation of the spacetime metric

\begin{equation}
\hat g_{ab}=\Omega^2(x) g_{ab},
\end{equation}
where $\Omega(x)$ is a non-vanishing smooth function, the spacetime coincidences [coordinates] are unchanged. This means that the spacetime measurements are invariant under (1.1). In other words, experimental observations [measurements] are unable to distinguish a metric with a hat from a metric without it.

Dicke has shown that the transformation (1.1) can be interpreted as a point-dependent transformation of the units of length, time and mass\cite{dk}. In effect, if one takes the arc-length as one's unit of measurement [length and time unit since the speed of light $c=1$] hence, since $d\hat s=\Omega(x) ds$, the units of length and time in the conformal frame are point-dependent even if $ds$ is a constant along a geodesic in the original frame. In the same way, in Brans-Dicke (BD) theory \cite{bdk} the mass unit transforms [under (1.1)] as $\hat m=\Omega^{-1}(x)m$. Hence, in the conformal frame the unit of mass $\hat m$ will be point-dependent even if $m$ [the unit of mass in the original formulation] is a constant.

In the classical papers \cite{dk,bdk} the authors put forth the requirement that the laws of physics must be invariant not only under general coordinate transformations but, also, under point-dependent transformations of the units of length, time and mass [these transformations are meaningless from the physical point of view]. In this paper I shall raise this last requirement to a cathegory of a postulate that I shall call as Brans-Dicke postulate: {\it The laws of physics, including the laws of gravitation, must be invariant under the transformations of the group of point-dependent transformations of the units of length, time and mass}.

Since there are decisive arguments against the assumption of a general conformal symmetry of the physical laws\cite{cgno},\footnotemark\footnotetext{The elegance and usefulness of this symmetry has been used, for instance, in Ref.\cite{dirac,canuto}} hence the usefulness and scope of the BD postulate should be discussed in detail. In this sense, in Sec. III of the present paper it will be shown that although, in the general case, conformal symmetry is not a symmetry of the laws of gravity [I mean those laws that are derivable from the generic low-energy actions studied here], there is a particular set of one-parameter conformal transformations of the metric tensor that constitute a group of symmetry of the conformal formulation of general relativity (conformal GR) [see below]. Therefore this group is identified with the one-parameter group of point-dependent transformations of the units of length, time, and mass and, consequently, conformal GR is acknowledged as the only viable [physically meaningful] formulation of the laws of gravity among those studied in the present paper.

However a class of conformal transformations of the metric remains that does not belong to this group of symmetry. It is, precisely, the class of conformal rescalings that are usually used to "jump" from one formulation of generic scalar-tensor, non-linear, and Kaluza-Klein theories of gravity [among others] to their conformal formulations. It is reminiscent of the fact that, in general, the laws of gravitation are not conformally invariant. In this sense I shall note that the geometrical interpretation of the conformal transformation (1.1) as a transformation of the units of length, time, and mass is not justified in the general case since invariance under the [physically meaningless] transformations of the units of measure is an obvious requirement any consistently formulated physical law must share. For this reason, in what follows, the conformal transformations that constitute a group of symmetry of conformal GR I shall call as "group of transformations of the units of length, time, and mass" or, simply, "transformations of units". Meanwhile, the transformations that allow jumping from the original fromulation of the theory to its conformal formulation I shall call as "proper conformal transformations" or, simply, "conformal transformations". Hence, the set of conformal rescalings of the metric [Eq.(1.1)] consists of the "transformations of units" and of the "conformal transformations". I remark that, in virtue of the discussion above, this distinction is absolutely neccessary. There are now clear the differences between the statements such as, for instance, "the laws of gravity are not invariant under the conformal rescalings of the metric" and "the laws of gravity must be invariant under the transformations of the units of measure". Both statements are correct. They are not alternative, but, complementary to each other.

In the present paper I shall study two generic effective theories of gravity and their conformal formulations. One of these generic effective theories is BD theory itself. It is based upon the action\cite{bdk}

\begin{equation}
S=\int d^4x \sqrt{-g}(\phi\;R-\frac{\omega}{\phi}(\nabla\phi)^2+16\pi L_{matter}),
\end{equation}
where $R$ is the curvature scalar, $\phi$ is the BD scalar field [the dilaton], $\omega$ is the BD coupling constant and $L_{matter}$ is the lagrangian of the matter fields that are minimally coupled to the metric. Under the change of variable $\phi\rightarrow\,e^\psi$, BD theory can be written in the string frame\cite{ps}

\begin{equation}
S_1=\int d^4x\sqrt{-g}\;e^\psi(R-\omega(\nabla\psi)^2+16\pi e^{-\psi}L_{matter}). 
\end{equation}                                    

Dicke used the conformal transformation (1.1) with

\begin{equation}
\Omega^2=e^\psi
\end{equation}
to rewrite the action (1.3) [or (1.2)] in the Einstein frame, i.e., in a frame where the dilaton is minimally coupled to the curvature\cite{dk}

\begin{equation}
S_2=\int d^4x \sqrt{-\hat g}(\hat R-(\omega+\frac{3}{2})(\hat\nabla\psi)^2+16\pi e^{-2\psi}L_{matter}),
\end{equation}
where $\hat R$ is the curvature scalar in terms of the metric with a hat and the matter fields are now non-minimally coupled to the dilaton $\psi$.

Another effective theory of gravity of BD-type was proposed by Magnano and Sokolowski\cite{ms}. They proposed to look at a BD-type theory with minimal coupling of the matter fields in the Einstein frame

\begin{equation}
S_3=\int d^4x\sqrt{-g}\;(R-\alpha(\nabla\psi)^2+16\pi\;L_{matter}),
\end{equation}
where $\alpha\equiv\omega+\frac{3}{2}$. We must acknowledge, however, that (1.6) is just the canonical action of general relativity with an extra scalar [dilaton] field. When $\alpha=0$ [$\psi$ arbitrary] or when $\psi=const$ we recover usual Einstein's formulation of GR.\footnotemark\footnotetext{Both cases with $\alpha=0$ [$\psi$ arbitrary] and with $\psi=const$ must be considered separatelly}

Under the conformal transfromation (1.1), (1.4), the action (1.6) can be written in the string frame

\begin{equation}
S_4=\int d^4x\sqrt{-\hat g}\;e^{-\psi}(\hat R-(\alpha-\frac{3}{2})(\hat\nabla\psi)^2+16\pi\;e^{-\psi}\;L_{matter}).
\end{equation}

The theory derivable from the action (1.7) I shall call as "conformal GR" or, alternatively, "string-frame GR". In Sec. III the BD postulate formulated at the beginning of this section will be used to select which of the actions $S_1$, $S_2$, $S_3$, and $S_4$ [if any] represents a physically meaningful formulation of the laws of gravity, i.e., a theory of gravity that is invariant under transformations of the units of measure.

A final remark on the viewpoint to be developed in the present paper. The requirement of invariance under point-dependent transformations of units leads one to consider geometries that admit units of measure that may vary along transport. These geometries are called generically as Weyl geometries\cite{weyl}. On the other hand, theories that are derivable from the actions $S_1$ and $S_3$ are {\it a priori} [and {\it a posteriori}] uniquely linked with manifolds of Riemannian structure\cite{novello}.  Therefore, their conformal formulations that are derivable from the actions $S_2$ and $S_4$ respectively must be uniquely linked with manifolds of conformally-Riemannian structure. Spacetimes of conformally-Riemannian structure are acknowledged also as Weyl-integrable spacetimes (WISTs)\cite{novello}.\footnotemark\footnotetext{This point will be discussed in detail in Sec. II} It is a special kind of Weyl spacetimes that are of particular interest since they are free of the "second clock effect" that leads to observational inconsistencies (see Ref.\cite{novello} and references therein). In Sec. III it will be shown that WIST configurations are invariant under the transformations of the one-parameter group of point-dependent transformations of the units of length, time, and mass. Hence, adopting the requirement of invariance under the transformations of the units of measure as a basic postulate leads one to consider Weyl-integrable geometries as real alternatives to Riemann spaces for the geometrical interpretation of the physical laws. Moreover, from the physical standpoint WIST configurations are preferred over Riemannian configurations. Other arguments pointing at this conclusion can be found in Ref.\cite{novello,vp,ja}.

The paper has been organized in the following way. In Sec. II a brief survey on the fundamentals of Weyl-integrable geometry is given and the linkage between matter couplings and the geometric structure of the underlying manifolds is shown. In Sec. III I study the one-parameter group of point-dependent transformations of the units of length, time, and mass and, I inspect the different generic effective actions $S_1$, $S_2$, $S_3$, and $S_4$ in respect to their invariance properties under the transformations of this group. The cosmological singularity problem is treated, in Sec. IV, within the context of string cosmology with non-minimal coupling of the dilaton to a perfect [barotropic] fluid of solitonic p-brane. The results obtained are to be geometrically interpreted on the grounds of a Weyl-integrable geometry. The implications of  these results for the Mach's principle are briefly outlined in Sec. V. Finally, in Sec. VI, some concluding remarks are given.

\section{Conformally-Riemannian geometries and matter couplings}

As it has been properly remarked in Sec. I, according to the canonical Einstein's formulation of general relativity [Action (1.6) with $\alpha=0$ or $\psi=const$] "...the structure of physical spacetime must correspond unequivocally to that of a Riemann manifold..."\cite{novello}. In general, theories with minimal coupling of the matter content to the metric field are naturally linked with manifolds of Riemann structure. In fact, in theories with the matter part of the action of the kind

\begin{equation}
S_{matter}=16\pi\int d^4x\sqrt{-g}\;L_{matter},
\end{equation}
the [time-like] matter particles follow free-motion paths that are solutions of the following differential equation 

\begin{equation}
\frac{d^2 x^a}{ds^2}+\{^{\;\;a}_{mn}\}\frac{dx^m}{ds}\frac{dx^n}{ds}=0,
\end{equation}
where $\{^{\;a}_{bc}\}\equiv\frac{1}{2}g^{an}(g_{bn,c}+g_{cn,b}-g_{bc,n})$ are the Christoffel symbols of the metric. Eq.(2.2) coincides with the equation defining [time-like] geodesic curves in Riemannian spacetimes. In general a Riemann configuration is characterized  by the requirement that the covariant derivatives of the metric tensor vanish, i.e.,

\begin{equation}
g_{ab;c}=0,
\end{equation}
where semicolon denotes covariant differentiation in a general affine sense.\footnotemark\footnotetext{We use, mainly, the notation of Ref.\cite{novello}} Fulfillment of this condition leads the manifold affine connections $\Gamma^a_{bc}$ to become identical to the Christoffel 
symbols $\{^{\;a}_{bc}\}$ of the metric (i.e., $\Gamma^a_{bc}\equiv\{^{\;a}_{bc}\}$). Hence, a Riemann configuration of spacetime is characterized by the  requirement that $g_{ab\|c}=0$, where the double bar denotes covariant differentiation defined through the Christoffel symbols of the metric [instead of the affine connections $\Gamma^a_{bc}$]. This requirement implies that vector lengths do not change under parallel transport, meaning that the units of measure of the geometry are point-independent.

Therefore, in both string-frame BD theory that is derivable from the action $S_1$ and Einstein frame general relativity that is derivable from $S_3$, the structure of the underlying manifold is Riemannian in nature since both effective actions show minimal coupling of the Lagrangian of the ordinary matter fields to the dilaton. Both theories are, therefore, compatible with a system of point-independent physical units.

Under the conformal rescaling (1.1) the actions $S_1$ and $S_3$ are mapped into their conformal actions $S_2$ and $S_4$ respectively but, at the same time, manifolds of Riemann structure are mapped onto manifolds of conformally-Riemannian structure. Therefore, in theories that are derivable from the actions $S_2$ (Einstein frame BD theory) and $S_4$ (string-frame GR) the underlying manifolds are of conformally-Riemannian nature. Conformally-Riemannian manifolds are also acknowledged as Weyl-integrable spaces\cite{novello}. In effect, under the rescaling (1.1), (1.4), the Riemannian requirement Eq. (2.3) is transformed into the following requirement

\begin{equation}
\hat g_{ab;c}=\psi_{,c}\;\hat g_{ab},
\end{equation}
where now semicolon denotes covariant differentiation in a general affine sense with $\hat\Gamma^a_{bc}$ being the affine connection of the conformal manifold. It is given through the Christoffel symbols of the metric with a hat $\{^{\;a}_{bc}\}_{hat}$ and the derivatives of the scalar [dilaton] function $\psi$ as

\begin{equation}
\hat\Gamma^a_{bc}\equiv\{^{\;a}_{bc}\}_{hat}-\frac{1}{2}(\psi_{,b}\;\delta^a_c+\psi_{,c}\;\delta^a_b-\hat g_{bc}\hat g^{an}\;\psi_{,n}).
\end{equation}

If one compares Eq. (2.4) with the requirement of nonvanishing covariant derivative of the metric tensor $\hat g_{ab}$ in the most generic cases of Weyl geometries\cite{novello}:

\begin{equation}
\hat g_{ab;c}=f_c\;\hat g_{ab},
\end{equation}
in which $f_a(x)$ is the Weyl gauge vector, one arrives at the conclusion that the conformally-Riemannian geometry -characterized by Eq. (2.4)- is a particular type of Weyl geometry in which the gauge vector is the gradient of a scalar function $\psi$ [the dilaton]. This particular type of Weyl spaces is called a Weyl-integrable spacetime (WIST) since length variations are integrable along closed paths: $\oint dl=0$, where $dl=l\;dx^n\psi_{,n}$, and $l\equiv\hat g_{nm} V^n V^m$ is the length of the vector $V^a(x)$ being parallelly transported along of the closed path. For this reason, in manifolds of WIST configuration, the disagreement with observations due to the "second clock effect"\cite{vp} -that is inherent to Weyl spacetimes in general- is overcome\cite{novello}. Therefore, under the transformation (1.1), (1.4) Riemann geometry is mapped into a WIST geometry. Hence, theories that are derivable from the actions $S_2$ and $S_4$ [that are conformal to $S_1$ and $S_3$ respectively] are naturally linked with manifolds of WIST configuration. In other words, in theories showing non-minimal coupling of the matter fields to the metric, in particular with the matter part of the action of the kind

\begin{equation}
S_{matter}=16\pi\int d^4x\sqrt{-\hat g}\;e^{-2\psi}\;L_{matter}
\end{equation}
[it is conformal to (2.1)], the nature of the underlying manifold is that of a WIST configuration. The equations of free-motion of a material test particle that are derivable from Eq. (2.7)

\begin{equation}
\frac{d^2 x^a}{d\hat s^2}+\{^{\;\;a}_{mn}\}_{hat}\frac{dx^m}{d\hat s}\frac{dx^n}{d\hat s}-\frac{\psi_{,n}}{2}(\frac{dx^n}{d\hat s}\frac{dx^a}{d\hat s}-\hat g^{na})=0,
\end{equation}
are conformal to Eq. (2.2). Equations (2.8) coincide with the equations defining geodesic curves in spacetimes of WIST configuration. These can be also obtained with the help of the variational principle $\delta\int e^{-\frac{\psi}{2}} d\hat s=0$, that is conformal to $\delta\int ds=0$.

The requirement of nonvanishing covariant derivative of the metric tensor $\hat g_{ab}$ [Eq. (2.4)] implies that vector lengths may vary along transport or, in other words, that the units of measure may change locally. Therefore WIST geometry represents a generalization of Riemann geometry to include units of measure with point-dependent length.

Summingup. Under the conformal rescaling (1.1), (1.4), theories with minimal coupling of the matter fields to the metric [for instance those with the matter part of the action of the kind (2.1)] being uniquely linked with manifolds of Riemann structure, are mapped into theories with non-minimal coupling of the matter fields to the metric [with the matter part of the action of the kind (2.7)] in which the nature of the underlying manifold is of WIST structure. In particular, the Riemannian geodesics of the metric $g_{ab}$ are mapped under (1.1), (1.4) onto geodesics in manifolds of a WIST structure that are specified by the conformal metric $\hat g_{ab}$ and the gauge vector $\psi_{,a}$.

\section{Group of the transformations of the units of length, time and mass}

Now we shall study the conformal transformation (1.1) with $\Omega(x)$ defined in the following way:

\begin{equation}
\Omega^2(x)=e^{\sigma\psi(x)},
\end{equation}
where $\sigma$ is some [arbitrary] constant parameter. Transformation (1.1), (3.1) is more general than (1.1), (1.4) [(1.4) is a particular case of (3.1) when $\sigma=1$].

As before, following Dicke, (1.1), (3.1) -being a conformal rescaling of the metric- may be interpreted geometrically as a one-parameter, point-dependent transformation of the units of length, time, and mass. A very interesting situation is obtained if one makes the change of variable

\begin{equation}
\hat\psi=(1-\sigma)\psi.
\end{equation}

Under the one-parameter set of transformations (1.1), (3.1), and (3.2), the basic requirement of a WIST geometry [Eq. (2.4)] is preserved, i.e., under these transformations, Eq. (2.4) is invariant in form: $g_{ab;c}=\hat\psi\;g_{ab}$. In particular the geodesic equation of WIST geometry [Eq. (2.8)] is invariant in form too under these transformations. Unfortunately it is not true for Riemann geometry; neither the basic requirement that $g_{ab\|c}=0$ nor the geodesic equation (2.2) is invariant under (1.1), (3.1), and (3.2).

It can be checked that the purely gravitational part of the actions $S_1$ and $S_4$ showing non-minmal coupling of the dilaton to the curvature are invariant in form under (1.1), (3.1), and (3.2), and the parameter transformation\footnotemark\footnotetext{A demonstration of this invariance for the purely gravitational part of the Jordan frame action of BD theory -with a little different notation and variable replacements- can be found in Ref.\cite{far}}

\begin{equation}
\hat\alpha=\frac{\alpha}{(1-\sigma)^2},
\end{equation}
where $\sigma\neq 1$. The purely gravitational part of the actions $S_2$ and $S_3$ is not invariant under the above transformations. The same is true for the matter action (2.1). On the contrary, the matter action (2.7) is invariant in form under (1.1), (3.1-3.3). In fact, under (1.1), (3.1) the Eq. (2.7) can be written as

\begin{equation}
S_{matter}=16\pi\int d^4x \sqrt{-g}\;e^{2(\sigma-1)\psi}\;L_{matter}.
\end{equation}

The demonstration is completed if we replace in Eq. (3.4) $\psi$ with the help of Eq. (3.2).

Therefore, if we check the actions $S_1$, $S_2$, $S_3$, and $S_4$ in respect to their invariance properties under the transformations (1.1), (3.1-3.3), the only surviving action is $S_4$. In other words, among the different generic theories of gravity studied here, the only one that is invariant under the above transformations is the conformal formulation of general relativity [the string-frame GR].

The set of transformations (1.1), (3.1-3.3) with $\sigma\neq 1$ is a one-parameter Abelian group of transformations\cite{far}. A composition of two successive transformations with parameters $\sigma_1$ and $\sigma_2$ yields a transformation of the same kind with parameter $\sigma_3=\sigma_1+\sigma_2-\sigma_1\sigma_2$. The identity of this group is the transformation with $\sigma=0$ [no transformation]. The inverse of the transformation with parameter $\sigma$ is a transformation with parameter $\bar\sigma=\frac{\sigma}{\sigma-1}$. Since the order of the composition is irrelevant the group is commutative.  

I call attention upon three important aspects of the transformations of this group. First, these are conformal transformations of the metric and, hence, may be geometrically interpreted as point-dependent transformations of the units of length, time, and mass. Second, there is a formulation of the laws of gravity -namely, conformal GR or string-frame GR- that is invariant under the elements of this group of transformations. Third, arguments [of both axiomatic and observational character] have been put foth in the literature that point at WIST geometries as the most viable framework for the geometrical interpretation of the physical laws and, consequently, for the confrontation of physical observations\cite{vp,ja}. Hence, I feel, it is not casual that just WIST geometries possess the above group of symmetry. For these reasons I shall identify this group with the one-parameter group of point-dependent transformations of the units of length, time, and mass. It is important to remark that the transformation (1.1), (1.4) [it corresponds to the particular case when in (3.1) $\sigma=1$] does not belong to this group and, therefore, it may not be interpreted properly as a transformation of units. It is just a transformation that allows jumping from one formulation of the theory to the conformal formulation.

If we use the BD postulate as a selection principle two crucial conclusions raise. First, the string-frame formulation of general relativity due to the action $S_4$ [Eq. (1.7)] is the only surviving [physically meaningful] formulation of the laws of gravity since, among those generic formulations studied here, it is the only one that is invariant under the transformations of the units of length, time, and mass. Second -and by the same reason- a WIST geometry seems to be the most adequate framework for the geometrical interpretation of the laws of gravity.

Some colleagues may detract these conclusions with the argument that canonical relativity [action $S_3$ with $\alpha=0$ or $\psi=const$] is mathematically consistent and describes all current observations with remarkable accuracy\cite{will}. However, I recall that arguments of observational character can not help us while choosing one or another conformal formulation of a given theory [as, for example, formulations due to $S_3$ and $S_4$] since physical observations [measurements] are unchanged by a conformal rescaling of the spacetime metric [Eq. (1.1)]. Hence, for instance, both theories derivable from actions $S_3$ and $S_4$ are equally consistent with the same experimental data. Besides, it may happen that a universe of WIST structure evolves into a universe of a Riemann configuration. So, arguments based on the current status of the observational evidence are irrelevant.

Since string theory seems to be the most serious candidate for a final theory of spacetime, in what follows I shall identify the theory derivable from the action $S_4$ with the low-energy [and low-curvature] limit of string theory. Consequently, the cosmological singularity problem will be treated within the context of string cosmology on Friedmann-Robertson-Walker (FRW) spacetimes of WIST configuration.

\section{Regular bouncing FRW cosmologies}

In this section I propose to study Friedmann-Robertson-Walker cosmology within the context of the string frame formulation of general relativity [Eq. (1.7)]. The geometrical interpretation of the results obtained is to be given on the grounds of WIST geometry. The physical metric is therefore the metric with a hat $\hat g_{ab}$.

FRW spacetimes of Riemannian structure are given by the line element

\begin{equation}
ds^2=-dt^2 + a^2 (\frac{dr^2}{1-kr^2}+r^2 d\Omega^2),
\end{equation}
where $a(t)$ is the time-dependent scale factor and $d\Omega^2=d\theta^2+\sin^2 \theta d\phi^2$. I shall concentrate on flat [$k=0$] and open [$k=-1$] FRW universes since the case with $k=+1$ requires a more detailed analysis. The FRW spacetime is suppossed to be filled with a perfect fluid with the barotropic equation of state $p=(\gamma-1)\mu$, where the barotropic index $0<\gamma<2$. The field equations derivable from the effective action (1.6) [canonical action of GR theory] can be reduced to the following equation for determining the Riemannian scale factor

\begin{equation}
(\frac{\dot a}{a})^2+\frac{k}{a^2}=\frac{M}{a^{3\gamma}}+\frac{\alpha N^2}{a^6},
\end{equation}
where the overdot means derivative with respect to the cosmic time $t$. $M$ and $N$ are arbitrary integration constants. For the dilaton one has

\begin{equation}
\psi^\pm = \psi_0 \mp \sqrt{6}N\;I(a),
\end{equation}
where $\psi_0$ is another integration constant and $I(a)$ is defined as

\begin{equation}
I(a)\equiv I(a,\gamma,\alpha,k)\equiv\int\frac{da}{\sqrt{Ma^{8-3\gamma}+\alpha N^2 a^2-k a^6}}.
\end{equation}

The choice of the "+" and "-" signs in Eq. (4.3) defines two different branches of the solution. The curvature scalar of the Riemannian FRW spacetime $R=\frac{3M}{a^6}((4-3\gamma)a^{3(2-\gamma)}-\frac{2\alpha N^2}{M})$ shows that there is a curvature singularity at $a=0$ that corresponds to the cosmological singularity at the beginning of the cosmic time $t=0$.

However, as it has been properly remarked in Sec. III, the string-frame effective action (1.7) [that is naturally coupled to a WIST geometry] is the only surviving candidate [among those studied here] for a physically meaningful formulation of the laws of gravity. For this reason I shall interested in the behaviour of the relevant magnitudes and relationships precisely in this frame [magnitudes with a hat]. The string-frame curvature scalar, for instance, can be written as a function of the Riemannian scale factor $a$,

\begin{equation}
\hat R^\pm=\frac{3Me^{-\psi_0}}{a^6}\;e^{\pm\sqrt{6}N\;I(a)}((4-3\gamma)a^{3(2-\gamma)}-(2\alpha-3)\frac{N^2}{M}).
\end{equation}

The string-frame proper time $\hat t$ is related with $t$ through $\hat t=\int\Omega dt$. This last integral can be transformed into an integral over the Riemannian scale factor $a$

\begin{equation}
\hat t^\pm=e^\frac{\psi_0}{2}\int e^{\mp\frac{\sqrt{6}}{2}N\;I(a)}\frac{da}{\dot a}.
\end{equation}

The occurrence of a spacetime singularity ca be traced out with the help of the Raychaudhuri equation as a function of the Riemannian scale factor

\begin{equation}
\dot{\hat \theta}^\pm=e^{-\frac{\psi_0}{2}}e^{\mp\frac{\sqrt{6}}{2}N\;I(a)}(-\frac{9\gamma M}{2 a^{3\gamma}}-\frac{9(\alpha+\frac{1}{2}) N^2}{a^6}+\frac{3k}{a^2}\pm\frac{6\sqrt{6}N}{a^3}\sqrt{\frac{M}{a^{3\gamma}}+\frac{\alpha N^2}{a^6}-\frac{k}{a^2}}), 
\end{equation}
where now the overdot means derivative with respect to the string-frame proper time $\hat t$ and $\hat\theta$ is the string-frame volume expansion of a congruence of fluid lines without vorticity and shear. In Eq. (4.7) I took the reversed sense of the proper time $\hat t$. In terms of the Riemannian scale factor this means that $a$ runs from infinity to zero.

In what follows I choose the "+" branch of the solution since it defines the physical branch of our solution, i.e., the branch with the desired evolution of the Newton's constant. In effect, in this case $G\sim e^\psi$ evolves from an infinite value in the infinite [gravity dominated] past -in terms of the string-frame proper time $\hat t$- into a finite constant value in the infinite future. If we choose the "+" sign in Eq. (4.7) hence the 4th term under brackets in the right-hand side (RHS) of this equation induces expansion of fluid lines instead of contraction. In particular, if the parameter $\alpha$ is in the range $0\leq\alpha\leq\frac{1}{6}$ [$-\frac{3}{2}\leq\omega\leq -\frac{4}{3}$], a detailed analysis of Eq. (4.7) shows that, in the reversed sense of the proper time $\hat t$ the fluid lines converge into the past until the scale factor $a$ becomes small enough. Then the 4th term in brackets in the RHS of Eq. (4.7) dominates over the others and, hence, further contraction of fluid lines is not allowed. Instead, the fluid lines begin to diverge for ever into the infinite past. Hence, bouncing universes are obtained if $0\leq\alpha\leq\frac{1}{6}$. These bouncing universes are regular everywhere in the range $0\leq a\leq\infty$ [$-\infty\leq\hat t\leq +\infty$]. In particular the curvature scalar [Eq. (4.5)] is bounded and well behaved everywhere in the above range. 

The vanishing of the cosmological singularity in flat and open FRW universes of WIST structure [in the string-frame formulation of general relativity] can be explained as due to the fact that the incomplete [into the past] fluid lines over spacetimes of Riemann configuration are mapped, under (1.1), (1.4) onto complete curves over their conformal Weyl-integrable spacetimes. Under this transformation singular universes of Riemannian structure emerge as regular bouncing universes of WIST configuration.

This result may be properly interpreted within the context of string cosmology. According to the viewpoint developed in the present paper, the only surviving candidate for a physically meaningful classical [low-energy] theory of spacetime is that due to the effective action $S_4$ [Eq. (1.7)]. Therefore, it is natural to identify the string-frame formulation of general relativity with the low-energy limit of string theory. Consequently I shall study a gas of solitonic matter -treated as a barotropic perfect fluid- living in an homogeneous and isotropic FRW spacetime of WIST configuration. It is important to note that, for the solitonic $p$-brane, the parameter $\alpha$ in Eq. (1.7) can be wirtten in the following form\cite{ps,rama}: $\alpha=\frac{1}{2}\frac{(p+1)^2}{p^2+3}$. Therefore, for the zero-brane [$p=0$] $\alpha=\frac{1}{6}$, while for the instanton [$p=-1$] $\alpha=0$.

In this case, as I have already shown, the string effective action Eq. (1.7) yields regular bouncing universes if the coupling parameter $\alpha$ is restricted to the range $0\leq\alpha\leq\frac{1}{6}$ [$-\frac{3}{2}\leq\omega\leq -\frac{4}{3}$]. This means that, if the Weyl-integrable spacetime is filled with a gas of fundamental string [$\alpha=\frac{1}{2}$ or $\omega=-1$], it will develop cosmological singularities while, if it is filled with a gas of solitonic $p$-branes [in particular 0-brane and/or instanton], it will behave as a regular bouncing spacetime. It is very encouraging, at this moment in the discussion, to call attention upon a derivation of a field theory mechanism where heavy states become light and resolve the moduli space singularities\cite{ssw}. According to a derivation of this mechanism, the solitonic $p$-branes are heavy when the string coupling is weak, but they become light when the coupling is strong\cite{pw,rama}. Therefore they are copiously produced and dominate the universe during the strong string coupling phase. 

In the theory developed in this paper [string cosmology of Weyl-integrable spacetimes], for the relevant branch of our solution [the "+" branch], the effective string couplig $g^+_s\sim e^{-I(a)}$, therefore it becomes strong as we approach the high-curvature regime [in Riemannian GR] with $a=0$.\footnotemark\footnotetext{When $a$ approaches zero $I(a)$ approaches $-\infty$ so $g^+_s$ approaches $+\infty$} Hence, in the limit $a\rightarrow 0$ [$\hat t\rightarrow -\infty$] the $p$-branes become light and are copiously produced. This mechanism explains why the solitonic degrees of freedom such like the 0-brane [$\alpha=\frac{1}{6}$] and the instanton [$\alpha=0$] dominate the WIST universe at early times $\hat t=-\infty$ and, consequently remove the cosmological singularity being a common feature of spacetimes of Riemannian structure.

\section{Global features of the universe and local physics}

In this section I shall briefly discuss the implications that the former considerations and results have for the Mach's principle. This last, being a phylosophical principle, deals with the way in which the global features of the universe influence the local behaviour of matter. As before I shall take the string-frame formulation of general relativity [Eq. (1.7)] -where the underlying manifold is of WIST structure- for addressing this question. Therefore the magnitudes with a hat will be the physical magnitudes. The inertial mass of some elementary particle, for instance, will be given by

\begin{equation}
\hat m=e^{-\frac{\psi}{2}} m,
\end{equation}
where $m$ is some constant parameter with dimension of mass. It serves as a local measure of the inertial properties of the elementary particles. At the same time the arc-length between two successive events along a geodesic curve is given by

\begin{equation}
d\hat s=e^\frac{\psi}{2} ds,
\end{equation}
where $ds$ is some constant spacetime separation parameter for a given succession of events. The arc-length between successive events along a geodesic curve serves for the characterization of the local metric properties of the spacetime.

I shall look at the local behaviour of the above magnitudes in the FRW universe studied in the former section. If we take into account the Eq. (4.3), hence we can write the following approximate expressions for the inertial mass of a particle

\begin{equation}
\hat m^\pm\sim e^{\pm\sqrt{6}N\;I(a)},
\end{equation}
and for the line element

\begin{equation}
d\hat s^\pm\sim e^{\mp\sqrt{6}N\;I(a)}.
\end{equation}

As before we shall take the "+" sign in Eqs. (5.3) and (5.4) since it defines the physical branch of our solution. From these equations one sees that both local inertial properties of the elementary particles and local metric properties of spacetime depend on the boundary conditions -that are specified by the values of the constants $N$, $M$ and $\psi_0$- and, in general, on the global properties of the universe such as the curvature $k$ of the FRW spacetime. In fact, the indefinite integral (4.4) depends on the parameters $\alpha$, $\gamma$, and $k$. Take, for instance, the most simple situation with $\alpha=0$ [$\omega=-\frac{3}{2}$]. In this case $I(a)\equiv I(a,\gamma,k)$ is a function of the barotropic index and of the curvature only. For a flat FRW universe we have

\begin{equation}
I(a)^\gamma\equiv I(a,\gamma,0)=-\frac{2\;a^{\frac{3}{2}(\gamma-2)}}{3\sqrt{M}(2-\gamma)}.
\end{equation}

For an open FRW universe we have, instead,

\begin{equation}
I(a)^\gamma\equiv I(a,\gamma,-1)=\frac{2}{2-3\gamma)\sqrt{M}}\int\frac{X^\frac{6-3\gamma}{3\gamma-2}dX}{\sqrt{X^2+\frac{1}{M}}},
\end{equation}
where we have introduced the variable $X\equiv a^\frac{2-3\gamma}{2}$. For a dust-filled universe [$\gamma=1$] we have

\begin{equation}
I(a)^1\equiv I(a,1,-1)=-\frac{2}{\sqrt{M}}(\frac{1}{3}(\frac{1}{a}+\frac{1}{M})^\frac{3}{2}-\frac{1}{M}\sqrt{\frac{1}{a}+\frac{1}{M}}),
\end{equation}
while for a radiation-filled universe [$\gamma=\frac{4}{3}$]

\begin{equation}
I(a)^\frac
{4}{3}\equiv I(a,\frac{4}{3},-1)=-\frac{1}{M}\sqrt{\frac{1}{a^2}+\frac{1}{M}}.
\end{equation}

 In all cases when $a$ approaches zero the function $I(a)$ approaches $-\infty$. Meanwhile, when $a$ approaches infinity $I(a)$ approaches $\frac{4}{3M}$ for a dust-filled [open] universe and $-\frac{1}{M}$ for a radiation-filled [open] universe. For a flat universe $I(a)$ approaches zero in this limit. This means that, in all studied cases, the inertial mass of an elementary particle evolves from being zero in the infinite past $\hat t=-\infty$ [$a=0$] to being a constant in the infinite future $\hat t=+\infty$ [$a=\infty$]. This last constant value depends on the fact whether the universe is flat or open. At the same time the metric properties of spacetime -that are given by the arc-length between successive [neighbouring] events along a geodesic curve- are undefined at the infinite past. In the infinite future the arc-length between two successive events tends to a constant finite value meaning that, in this limit, the universe approaches a state of Riemannian configuration.

This result admits the following explanation that is compatible with the Mach's principle. In the infinite past the universe was, supposedly, in the state of quantum vacum. The absense of ordinary matter in the universe in that state means that the local inertial properties of the elementary particles were undefined. As a result of the evolution of the universe ordinary matter is created at some stage of the evolution. In the infinite future, although the scale factor becomes infinite [meaning that each piece of matter is infinitely distant from each other, i.e., that the mean density of matter in the universe is zero], a large amount of matter is distributed over the spacetime. This global distribution of matter over the universe [according to the Mach's principle] influences the local inertial properties of the matter so, each elementary piece of matter is characterized by some [constant] value of its inertial mass. Meanwhile, in the infinite [quatum-dominated] past, the metric properties of spacetime were undefined by the same reason: the absense of ordinary matter in that state.\footnotemark\footnotetext{It may be that the same concept of spacetime losses its meaning at this quantum stage in the evolution of the universe}

\section{Conclusions} 

If one raises to a cathegory of a postulate the requirement that the physical laws [including the laws of gravitation] must be invariant under the transformations of the units of length, time, and mass, one is led to an important result: Only theories with non-minimal coupling of the dilaton both to the curvature and to the Lagrangian of the ordinary matter fields have chance of succes. This means that the experimental observations must be interpreted within the context of Weyl-integrable spacetimes, i.e., WIST configurations are preferred from the physical standpoint. Other arguments have been put forth in the bibliography that hint precisely at this result (see Ref.\cite{novello,vp,ja}).

In this context it seems likely that the low-energy effective action of string theory should reflect non-minimal coupling of the dilaton to the matter content. Consequently the string-frame formulation of general relativity [Eq. (1.7)] may be identified with the low-energy limit of string theory. This formulation of GR is observationally indistinguishable from the original [canonical] formulation of this theory. This hints at general relativity as the low-energy limit of string theory [I recall that usually theories of Brans-Dicke type are studied as such a limit of string theory\cite{ps,rama}].

A string cosmology with a perfect fluid of solitonic $p$-brane non-minimally coupled to the dilaton [the underlying manifold is of WIST structure] shows that the cosmological singularity is avoided. In fact, when the parameter $\alpha$ is in the range $0\leq\alpha\leq\frac{1}{6}$ one obtains Weyl-integrable FRW bouncing universes that are regular everywhere. On the other hand, according to a field-theory mechanism studied in Ref.\cite{rama}, at early times [$\hat t=-\infty$] the string coupling becomes strong, i.e., the solitonic degrees of freedom such like the 0-brane [$\alpha=\frac{1}{6}$] and the instanton [$\alpha=0$] become ligth and are therefore copiously produced. This mechanism explains why these solitonic degrees of freedom dominate the WIST universe at early times and, consequently, remove the cosmological singularity.

The experimental constrain $\omega>500$ [$\alpha>500$]\cite{will} does not arise in this case since canonical GR [Eq. (1.6) with $\alpha=0$, i.e., $\omega=-\frac{3}{2}$] describes all current observations with remarkable accuracy. The question is whether or not the values in the range $0\leq\alpha\leq\frac{1}{6}$ [$-\frac{3}{2}\leq\omega\leq -\frac{4}{3}$] fit too within the accuracy of present day experiments.

The Mach's principle finds a very desirable realization within the context of conformal general relativity in which the underlying manifold shows a WIST configuration. In this sense global features of the universe such as the spatial curvature $k$ [that is linked with the density of the matter distribution over the entire universe] influence the local physics, in particular, the local inertial properties of elementary particles [that are given through their rest masses] and the local metric properties of spacetime.

\begin{center}
{\bf ACKNOWLEDGMENT}
\end{center}

We acknowledge MES of Cuba by financial support of this research.


\begin{thebibliography}{99}

\bibitem{fgn} V. Faraoni, E. Gunzig and P. Nardone, Fundamentals of Cosmic 
Physics \textbf{20}(1999)121.
\bibitem{dk} R. H. Dicke, Phys. Rev. \textbf{125}(1962)2163.
\bibitem{bdk} C. Brans and R. H. Dicke, Phys. Rev. \textbf{124}(1961)925.
\bibitem{cgno} V. Canuto and J. Goldman, Nature\textbf{304}(1983)311; M. Novello and L. A. R. Oliveira, Int. J. Mod. Phys. A\textbf{1}(1986)943. 
\bibitem{dirac} P. A. M. Dirac, Proc. Roy. Soc. London A\textbf{165}(1938)199.
\bibitem{canuto} V. Canuto, P. J. Adams, S. H. Hsieh and E. Tsiang, Phys. Rev. D\textbf{16}(1977)1643.
\bibitem{ps} C. Park and S-J. Sin, Phys. Rev. D \textbf{57}(1998)4620.
\bibitem{ms} G. Magnano and L. M. Sokolowski, Phys. Rev. D \textbf{50}, 5039(1994); G. Magnano, 'Talk given at the XII Italian conference on general relativity and gravitation', Trieste, Sep. 26-30, 1994, gr-qc/9511027; L. M. Sokolowski in 'Proceedings of the 14th International Conference on General Relativity and Gravitation', Firenze, Italy 1995, M. Francaviglia, G. Longhi, L. Lusanna, E. Sorace (eds.), (World Scientific, 1997)337.
\bibitem{weyl} H. Weyl, Space, Time and Matter (Dover, NY, 1952).
\bibitem{novello} M. Novello, L. A. R. Oliveira and J. M. Salim, Int. J. Mod. Phys. D, Vol. 1, Nos 3 \& 4 (1993)641.
\bibitem{vp} V. Perlick, Class. Quantum Grav. \textbf{8}(1991)1369.
\bibitem{ja} J. Audretsch, Phys. Rev. D\textbf{27}(1983)2872.
\bibitem{far} V. Faraoni, Phys. Lett. A \textbf{245}(1998)26.
\bibitem{will} C. M. Will, Theory and experiment in gravitational physics(Cambridge University Press, 1993).
\bibitem{ssw} N. Seiberg and E. Witten, Nucl. Phys. B \textbf{426}(1994)19, ibid \textbf{472}(1996)349; A. Strominger, Nucl. Phys. B \textbf{451}(1995)109; H. Ooguri and C. Vafa, Phys. Rev. Lett. \textbf{77}(1996)3296.
\bibitem{pw} J. Polchinski, Phys. Rev. Lett. \textbf{75}(1995)4724; E. Witten, Nucl. Phys. B \textbf{460}(1996)335; ibid \textbf{443}(1995)85.
\bibitem{rama} K. Rama, Phys. Lett. B \textbf{408}(1997)91.


\end{thebibliography}
\end{document}